\documentclass[twoside,epsfig]{article}
\usepackage{fleqn,espcrc2,epsfig}

\usepackage{graphicx}
\usepackage[figuresright]{rotating}

\newcommand{\AmS}{{\protect\the\textfont2
  A\kern-.1667em\lower.5ex\hbox{M}\kern-.125emS}}

\title{Higgs Physics at a $\gamma \gamma$-Collider}

\author{Michael Melles\address{Department of Physics,\\
        Durham University, \\ 
        Durham, DH1 3LE, U.K.}%
        \thanks{Research supported by the EU Fourth Framework Programme 
                `Training and Mobility of Researchers' through a Marie Curie
		Fellowship.}}
\begin{document}

\begin{abstract}
High precision measurements of electroweak observables at $e^\pm$ colliders
indicate the existence
of a light Higgs boson below the $W^\pm$ threshold. If such a fundamental scalar
should be found in the near future it is important to fully investigate the
electroweak symmetry braking sector of the Standard Model. This is particularly
important for an intermediate mass Higgs as its existence might indicate
physics beyond the Standard Model, for instance in form of its minimal 
supersymmetric extension (MSSM).
In this work we
present first results on the expected precision of the partial 
$\Gamma ( H \longrightarrow \gamma \gamma )$ width at the $\gamma \gamma$ option
of a future linear collider. This quantity is sensitive to new physics as
heavy particles do not decouple in general and differences between the SM and
MSSM predictions can differ by up to 10\% even in the decoupling limit
of large pseudoscalar Higgs masses. This regime
is difficult and for some values of $\tan \beta$ impossible to  cover at the 
LHC. We find that the well understood background
process $\gamma \gamma \longrightarrow q \overline{q}$ allows for a ${\cal O}
(2\%)$ determination of $\Gamma ( H \longrightarrow \gamma \gamma )$ using
conservative collider parameters.
\vspace{1pc}
\end{abstract}

\maketitle

\section{Introduction}

At a future linear $e^\pm$ collider the possibility of using Compton
backscattered photons off the highly energetic and polarized incident
electron beams \cite{g,t} can be used for many interesting physics applications.
Important examples include the study of anomalous vector boson
couplings through the large $W^\pm$ cross section, the search
for heavy Higgs bosons up to $0.8 \sqrt{s}$ (compared to $0.5 \sqrt{s}$
in the $e^\pm$ mode), the CP-properties of fundamental scalars and
the partial $\Gamma (H \longrightarrow \gamma \gamma)$ width (and thus
the total Higgs-width assuming a knowledge of BR$( H \longrightarrow
\gamma \gamma )$ from the LHC for instance). In addition one has the
unique chance to study the polarized photon structure, 
complimentary processes involving supersymmetric particles (if they exist)
at comparable event rates and, more exoticly, can even study 
theories predicting new dimensions at the TeV scale through Kaluzza-Klein
excitations.

The main physics motivation of a future linear collider is not as much
the discovery
of new physics but its clarification. In analogous ways as SLC/LEP results
have made precision tests of the SM after the discovery of massive vector
bosons in hadronic collisions, the principle motivation for a linear collider
is based on the hope that precise measurements of processes at the 
electroweak scale can shed light on the deeper structure of physical laws
up to much higher energies, possibly GUT scales or evens regimes where
the gravitational coupling becomes comparable to the other gauge couplings.

\begin{center}
\begin{figure}[t]
\epsfig{file=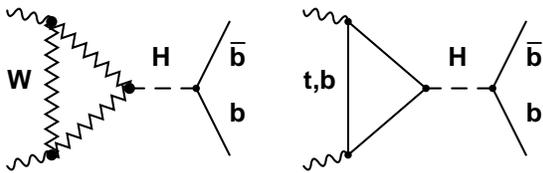,width=7.5cm} \vspace*{-1.5cm} \\
\caption{The Standard Model process $\gamma \gamma \longrightarrow H \longrightarrow
b \overline{b}$ is mediated by  $W-$boson and  $t-$ and $b-$quark loops.
}
\label{fig:ggH} $ $\vspace*{-1.7cm} \\
\end{figure}
\end{center}

In this context the partial $\Gamma ( H \longrightarrow \gamma \gamma )$ stands
out as an observable of considerable theoretical and experimental importance.
For intermediate mass Higgs bosons below 140 GeV it is the only decay channel
accessible at the LHC as a relatively background free process \cite{s}. 
At the Compton backscattered $\gamma \gamma$ option of the linear $e^\pm$
collider it represents the production process of a fundamental scalar.
All
charged particles contribute in the loop and heavy states
which obtain their mass through the Higgs mechanism
don't decouple due to the Yukawa coupling
at the vertex. 
Fig. \ref{fig:ggH} depicts the SM contributions from $W^\pm$ bosons and
t-and b-quarks. In 2-Higgs Doublet Models (2HDM), such as in the Higgs sector
of the MSSM, there are five fundamental scalars. The additional supersymmetry
of the MSSM leads to only two free parameters of the electroweak braking
sector, commonly chosen as the ratio of the vacuum expectation values of the
up- and down-type Higgs bosons, $\tan \beta$, and the mass of the pseudoscalar
Higgs boson, $m_A$. At tree level, all other masses and mixing angles are then
fixed. The diphoton partial Higgs width prediction of the MSSM depends
sensitively on these two parameters, even in the decoupling limit of large
$m_A > 500$ GeV. This regime, down to even lower $m_A$ for intermediate values
of $\tan \beta$, is not covered at the LHC \cite{s}.

It is therefore important at this stage of the design of future collider
schemes to investigate what the Compton collider option can contribute to the
high precision measurements of the parameters
of physics above the electroweak braking scales.
In this context we summarize the status of the background (BG) process
$\gamma \gamma \longrightarrow q \overline{q}$, $q = \{ b, c \}$ relevant to the
accurate determination of the signal (S) process $\gamma \gamma 
\longrightarrow H \longrightarrow b \overline{b}$ in the next
section. We then discuss our jet
definition and assumptions about the collider parameters and present 
first Monte Carlo results.
\vspace*{-1.5cm} \\

\section{Radiative Corrections}

The QCD corrections for both S and BG reactions are considerable and must be
taken into account properly in order to precisely determine the diphoton
Higgs width. In the next two sections we briefly summarize the radiative
corrections relevant to our desired high ($\sim$ few \%) accuracy for both
processes.

\subsection{The process $\gamma \gamma \longrightarrow q \overline{q}$}

In this section we review the QCD corrections to the continuum heavy quark
production in polarized photon-photon collisions. For the two possible
$J_z$ states we have at the Born level:
\begin{eqnarray}
\frac{d \sigma ( \gamma \gamma (J_z=0) \longrightarrow q \overline{q})}{d \; \cos
\; \theta} &=& \nonumber \\ \frac{12 \pi \alpha^2 Q^4_q \beta}{s \; ( 1 - \beta^2 \cos^2 \theta
)^2} (1 \!\!\!\!\!&-&\!\!\!\!\! \beta^4)  \label{eq:BJ0} \\
\frac{d \sigma ( \gamma \gamma (J_z=\pm 2) \longrightarrow q \overline{q})}{d \;
 \cos
 \; \theta} &=& \nonumber \\ \frac{12 \pi \alpha^2 Q^4_q \beta^3}{s \; ( 1 - \beta^2 \cos^2 
\theta
 )^2}(1\!-\! \cos^2 \theta)(2 \!\!\!\!\!\!&-&\!\!\!\!\!\! \beta^2 (1\!-\!\cos^2 \theta)) \label{eq:BJ2}
 \end{eqnarray}
where $\beta=\sqrt{1-4 m^2_q / s}$ denotes the quark velocity, $\sqrt{s}\equiv w
$
the $\gamma\gamma$ c.m.
collision energy, $\alpha^{-1}
\approx 128$ the electromagnetic coupling, $Q_q$ the charge of quark $q$ and
$m_q$ its pole mass. The  scattering angle of the produced (anti)quark
relative to the beam direction
is denoted by $\theta$. Eq.~\ref{eq:BJ0} clearly
displays the important feature that the $J_z=0$ cross section has a relative
$\frac{m^2_q}{s}$ suppression compared to  the $J_z=\pm 2$ cross
section. Since the Higgs $\gamma\gamma \to H \to q  \overline{q}$
process only occurs for $J_z = 0$, it is this polarization that is crucial
for the precision measurement of the Higgs partial decay width.

There are, however, important radiative corrections to the $J_z=0$ cross section as
large non-Sudakov double logarithm enter into loop effects \cite{jt} and real
Bremsstrahlung corrections can even remove the mass suppression \cite{bkos}.
In this work we will include the exact one loop corrections of Ref. \cite{jt}
(for both spin configurations) and use the recently achieved all orders
resummed DL-result of Ref. \cite{ms} for $J_z=0$:
\begin{eqnarray}
\sigma^{DL}_{{\rm virt}+{\rm soft}} \!\!\!&=&\!\!\! \sigma_{\rm Born}
\left\{ 1 +
{\cal F} \;\;
_2F_2 (1,1;2,\frac{3}{2}; \frac{1}{2}
{\cal F} )  \right. \nonumber \\ \!\!\!&&\!\!\! \left. +
2 \; {\cal F} \;\;
_2F_2 (1,1;2,\frac{3}{2}; \frac{C_A}{4 C_F}
{\cal F} ) \right\}^2 \times \nonumber \\
\!\!\!&&\!\!\! \exp \left( \frac{ \alpha_s C_F}{\pi} \left[ \log \frac{s}{m_q^2} \left(
\frac{1}{2} - \log \frac{s}{4 l_c^2} \right) \right. \right. \nonumber \\ 
\!\!\!&&\!\!\! \left. \left.
+ \log \frac{s}{4 l_c^2} -1 +
\frac{\pi^2}{3} \right] \right) \label{eq:vps}
\end{eqnarray}
where ${\cal F} = - C_F \frac{\alpha_s}{4 \pi} \log^2 \frac{s}{m_q^2}$
is the one-loop hard form factor,  $\alpha_s$ is taken as a fixed parameter,
and $l_c \ll \sqrt{s}$ is the soft-gluon upper energy limit.

In Ref.~\cite{ms2} it was pointed out that one needs to include at least
{\it four} loops (at  the cross section level) of the non-Sudakov
hard logarithms in order
to achieve positivity and stability. At this level of approximation there is an
additional major source of uncertainty in
the scale choice of the QCD coupling, two possible `natural' choices ---
$\alpha_s(m_H^2)$ and $\alpha_s(m_q^2)$  ---
yielding very different numerical results. These uncertainties were removed
in Ref. \cite{ms3} by employing a running coupling into each loop integration.
The correct running scale was found to be ${\bf l_\perp^2}$ in terms of the
perpendicular Sudakov components and was implemented by using 
\begin{eqnarray}
\alpha_s ({\bf l^2_\perp})
\!\!\!\!&=&\!\!\!\! \frac{\alpha_s(m^2)}{1\!+\! \left( \beta_0 \frac{\alpha_s(m^2)
}{\pi} +\beta_1 \left( \frac{\alpha_s(m^2)}
{\pi} \right)^2 \right) \log \frac{{\bf l^2_\perp}}{m^2}} \nonumber \\ 
\!\!\!\!&\equiv&\!\!\!\! \frac{\alpha_s(m^2)}{1+c \; \log \frac{{\bf l^2_\perp}}{m^2}} \label{eq:rc}
\end{eqnarray}
where $\beta_0=\frac{11}{12} C_A -\frac{4}{12} T_F n_F$,
$\beta_1 = \frac{17}{24} C_A^2 - \frac{5}{12} C_A T_F n_F- \frac{1}{4}
C_F T_F n_F$ and for QCD we have
$C_A=3$, $C_F=\frac{4}{3}$ and $T_F=\frac{1}{2}$ as usual.
Up to two loops the massless $\beta$-function is independent of the chosen
renormalization scheme and is gauge invariant in minimally subtracted schemes
to all orders. These features also hold for the
renormalization group (RG) improved form factors below and lead to \cite{ms3}:
\begin{equation}
\frac{ \sigma^{DL}_{RG}}{\sigma_{\rm Born}} = \left\{ 1 + \widetilde{{\cal F}}_h
^{RG}
\right\}^2 \exp \left( \widetilde{{\cal F}}^{RG}_{S_R} + 2 \widetilde{{\cal F}}^
{RG}_{S_V}
\right) \label{eq:srg}
\end{equation}
where the RG-massive Sudakov form factor is given by:
\begin{eqnarray}
\widetilde{{\cal F}}^{RG}_{S_R}+2 \widetilde{{\cal F}}^{RG}_{S_V}\!\!\!\!&=&\!\!\!\!
\frac{\alpha_s(m^2) C_F}{\pi} \left\{ \frac{1}{c}
\int_\frac{2l_cm^2}{(s+m^2) \sqrt{s}}^
\frac{2l_c\sqrt{s}}{s+m^2} \frac{d \beta_1}{\beta_1} \times \right. \nonumber \\ 
\!\!\!\!&&\!\!\!\! \log \frac{ 1+c \; \log
\left( \left(\frac{2l_c}{\sqrt{s}} -\beta_1 \right) \beta_1 \frac{s}{m^2}
\right)}{\left(
1+c \; \log \frac{s \beta_1}{m^2} \right)} \nonumber \\
\!\!\!\!&&\!\!\!\! - \frac{1}{c} \log \frac{s}{m^2} \log \frac{\alpha_s(2l_c \sqrt{s})}{
\alpha_s(s)} \nonumber \\  \!\!\!\!&&\!\!\!\!
- \frac{1}{c}\log \frac{2l_c}{\sqrt{s}} \log \frac{\alpha_s (2 l_c
\sqrt{s})}
{ \alpha_s \left( \frac{2l_cm^2}{\sqrt{s}}\right)} \nonumber \\ \!\!\!\!&&\!\!\!\!  -
\frac{1}{c^2} \log \frac{\alpha_s(m^2)\alpha_s(2l_c \sqrt{s})}{
\alpha_s(s) \alpha_s \left(\frac{2l_c m^2}{\sqrt{s}} \right)} \nonumber \\ \!\!\!\!&&\!\!\!\! \left. 
+ \!\frac{1}{2} \log \frac{s}{m^2} \!+\! \log \frac{s}{4 l_c^2} \!-\!1 \!+\! \frac{\pi^2}{3} \!
\right\}
\label{eq:RpVRGex}
\end{eqnarray}
\begin{figure}[t] $ $ \vspace*{-0.7cm} \\
\epsfig{file=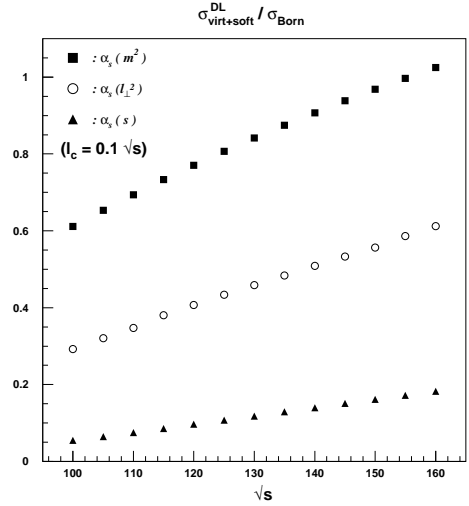,width=7.5cm} \vspace*{-2.0cm} \\
\caption{The effect of the renormalization group improved form factor (circles)
of Eq.~\ref{eq:srg}
in comparison to using the DL form factors of Eq.~\ref{eq:vps} with the indicated
values of the strong coupling. The energy cutoff is chosen to be $l_c=0.1 \sqrt{s}$.
The effect is displayed for the bottom quark
with $m_b=4.5$ GeV.}
\label{fig:vps} $ $ \vspace*{-1.7cm} \\
\end{figure}
and the hard (non-Sudakov) RG form factor by:
\begin{eqnarray}
\widetilde{{\cal F}}_h^{RG}
&=&\sum^\infty_{i=0} \int^s_{m^2} \frac{ d {\bf l^2_\perp}}{{\bf l^2_\perp}}
\left(\frac{C_F
}{2 \pi} \right)^{i+1} \left(
\frac{\alpha_s(m^2)}{ c}\right)^i \times \nonumber \\ &&
\frac{\alpha_s({\bf l^2_\perp})}{(i+1)!}
\log^{i+1} \frac{{\bf l^2_\perp}}{s} \log^i
\frac{\alpha_s(m^2)}{\alpha_s ( {\bf l^2_\perp} )}
+  \nonumber \\
&&2 \sum^\infty_{i=0} \int^s_{m^2} \frac{ d {\bf l^2_\perp}}{{\bf l^2_\perp}}
\frac{C_F C_A^i
}{2^{2i+1} \pi^{i+1}} \left(
\frac{\alpha_s(m^2)}{c} \right)^i \times \nonumber \\ &&
\frac{\alpha_s({\bf l^2_\perp})}{(i+1)!}
\log^{i+1} \frac{{\bf l^2_\perp}}{s} \log^i
\frac{\alpha_s(m^2)}{\alpha_s ( {\bf l^2_\perp} )}
\label{eq:rgff}
\end{eqnarray}
The effect of using the RG-form factors compared to the DL resummed result is
displayed in Fig. \ref{fig:vps}.
Here we choose the gluon energy cut $l_c=0.1 \sqrt{s}$. At one loop there is
no $l_c$ dependence while at higher orders there inevitably is.

\subsection{The process $\gamma \gamma \longrightarrow H \longrightarrow b
\overline{b}$}

An intermediate mass Higgs boson has a very narrow total decay width. It is 
therefore
appropriate to compare the total number of Higgs signal events with the number
of (continuum) background events integrated over a narrow energy window around
the Higgs mass. The size of this window depends on the level of monochromaticity
that can be achieved for the polarized photon beams.

\noindent In general, the number of events for the  (signal) process $S$
is given by
\begin{equation}
N_{S} = \int \frac{dL}{d w} \sigma_S(w) dw
\end{equation}
where $w$ denotes the center of mass energy. For $S \equiv \gamma \gamma 
\longrightarrow
H \longrightarrow b \overline{b}$ we have the following Breit-Wigner cross
section \cite{bkos}
\begin{equation}
\sigma_S(w) = \frac{16 \pi \Gamma (H \longrightarrow \gamma \gamma) \Gamma ( H
\longrightarrow b \overline{b})}{(w^2-m^2_H)^2+\Gamma_H^2 m^2_H} (\hbar c)^2
\end{equation}
where the conversion factor $(\hbar c)^2=3.8937966\times 10^{11}$~fb~GeV$^2$.
In the narrow width approximation we then find for the expected number of events
\begin{equation}
\frac{N_S}{(\bar{h}c)^2}\!\!=\!\! \left. \frac{d L_{\gamma \gamma}}{dw} \right|_{m_H} 
\!\!\!\!\!\!\!\!\frac{8 \pi^2 \Gamma (
H \!\!\longrightarrow \gamma \gamma) BR ( H \!\!\longrightarrow b \overline{b})}{m^2_H}
\end{equation}
To quantify this, we take the design parameters of the proposed TESLA
linear collider  \cite{tp}, which correspond to an integrated peak
$\gamma \gamma$-luminosity of 15 fb$^{-1}$ for the low energy running of the
Compton collider. The polarizations of the incident electron beams and the
laser photons are chosen such that the product of the helicities $\lambda_e
\lambda_{\gamma} = -1$.
This ensures high monochromaticity and polarization of the photon beams.
In this scenario we find a typical Higgs mass resolution of 10~GeV, so that for
the background process $BG \equiv \gamma \gamma \longrightarrow q \overline{q}$
we use
\begin{equation}
\frac{L_{\gamma \gamma}}{10\; {\rm GeV}} = \left. \frac{d L_{\gamma \gamma}}{dw}
\right|_{m_H}
\end{equation}
with $\left. \frac{d L_{\gamma \gamma}}{dw} \right|_{m_H}=$0.5 fb$^{-1}$/GeV.
The number of background events is then given by
\begin{equation}
N_{BG} = L_{\gamma \gamma} \sigma_{BG}
\end{equation}
In other words,
the number of signal events is proportional to $N_S \sim \left. \frac{d
L_{\gamma \gamma}}{dw} \right|_{m_H}$ while the number of  continuum heavy
quark production events is proportional to $N_{BG} \sim L_{\gamma \gamma}$.
In principle it is possible to use the exact Compton profile of the backscattered
photons to obtain the full luminosity distributions. The number of expected
events is then given as a convolution of the energy dependent luminosity and
the cross sections. Our approach described above corresponds to an effective
description of these convolutions, since these functions are not precisely
known at present. Note that the  functional forms currently used
generally assume that only one scattering takes place for each photon, which may
not
be realistic.  Once the exact luminosity functions
are experimentally determined  it is of course trivial to
incorporate them into a Monte Carlo program containing the physics described
in this paper.

\begin{center}
\begin{figure}[t]
\centering
\epsfig{file=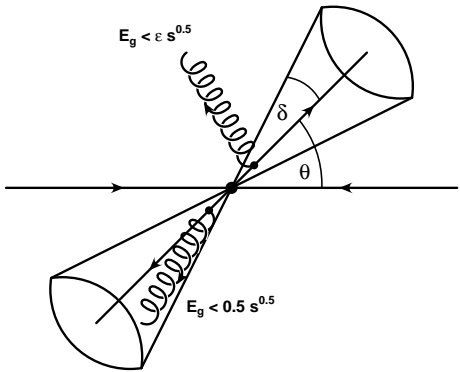,width=6.0cm} \vspace*{-1.0cm} \\
\caption{The parameters of the Sterman-Weinberg two jet definition used in this
work. Inside an angular cone of size $\delta$ arbitrary hard gluon Bremsstrahlung
is included. Radiation outside this cone is only permitted if the gluon energy
is below a certain fraction ($\epsilon$) of the incident center of mass energy.
The thrust angle is denoted by $\theta$.}
\label{fig:jet} $ $ \vspace*{-1.0cm} \\
\end{figure}
\end{center}

We next summarize the radiative corrections entering into the
calculation of the expected number of
Higgs events.
For the quantity $\Gamma (H \longrightarrow \gamma \gamma)$ there are three main
Standard Model contributions, depicted in Fig. \ref{fig:ggH}: the $W^\pm$
and $t-$ and $b-$quark loops. We include these at the one loop level, since the
radiative corrections are significant  only for
the $b-$quark.
The branching
ratio $BR(H \longrightarrow b \overline{b})$ is treated in the following way.
The first component consists of the partial $\Gamma (H \longrightarrow b 
\overline{b}
)$ width. Obviously we must use the same two-jet criterion
for the signal as for the background. For our purposes a cone-type algorithm is
most
suitable, and so we use the Sterman-Weinberg two-jet definition
depicted in   Fig.~\ref{fig:jet}. Note that the signal cross section is corrected
by the {\sl same} resummed renormalization group improved form factor
given in Eq.~\ref{eq:RpVRGex}, since this factor does not depend on the spin of
the
particle coupling to the final state quark anti-quark pair.
\vspace*{-2.1cm} \\
\begin{center}
\begin{figure}[t] $ $ \vspace*{-1.0cm} \\
\centering
\epsfig{file=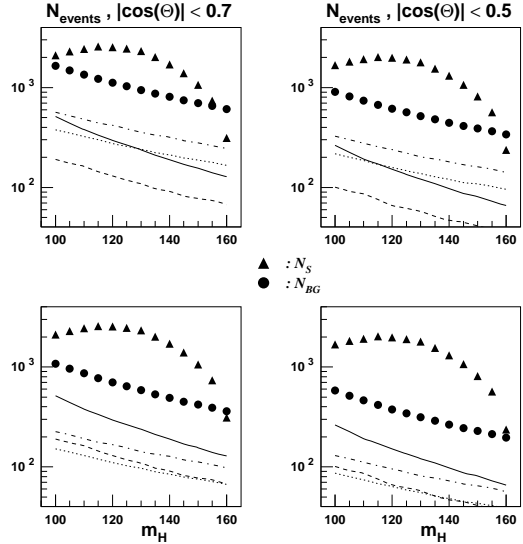,width=8.5cm} \vspace*{-1.8cm} \\
\caption{The number of both signal and background events for jet
parameters $\epsilon=0.1$ and $\delta=20^o$ and the indicated values
of the thrust angle $\theta$. The upper row assumes a ratio of
$J_0/J_2=20$ and the lower row of $50$. The background is composed of
bottom and charm contributions assuming 70 \% double b-tagging efficiency
and a 3 \% probability to count a $c \overline{c}$ pair as $b \overline{b}$.
The dash-dot line corresponds to $J_z=\pm2$ for $m_c$, the full line to
$J_z=0$ for $m_b$, the dotted line to $J_z=\pm2$ for $m_b$ and the dashed
line to $J_z=0$ for $m_c$. All lines are are normalized to add up to the
total background and all radiative corrections discussed in the
text are included.}
\label{fig:ne203} $ $ \vspace*{-1.25cm} \\
\end{figure}
\end{center}
In addition we use the exact one-loop corrections from Ref.~\cite{bl}. These
revealed that
the largest radiative corrections are well described by using the running quark
mass evaluated at the Higgs mass scale. We therefore resum the leading running
mass term to all orders.
For the real Bremsstrahlung corrections we use our own $H \longrightarrow
q \bar q g$
matrix elements. An important
check is obtained by integrating over all phase space and reproducing the analytical
results of Ref.~\cite{bl}.

The second quantity entering the branching ratio is the total Higgs width $\Gamma_H$.
Here we use the known results summarized in Ref.~\cite{s}, and include
the partial Higgs to $b \overline{b}$, $c \overline{c}$, $\tau^\pm$, $WW^*$,
$ZZ^*$ and $gg$ decay widths with all relevant radiative corrections.

\vspace*{-0.3cm} 
\section{Numerical Results}

\begin{center}
\begin{figure}[t] $ $ \vspace*{-1cm} \\
\centering
\epsfig{file=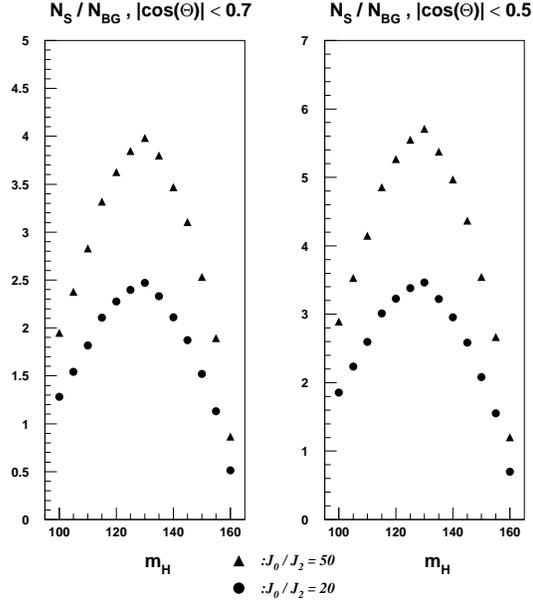,width=8.5cm} \vspace*{-1.3cm} \\
\caption{The ratio of signal to background events based on the jet parameters
of Fig. \ref{fig:ne203}. The smaller phase space cut $| \cos \theta |<0.5$ gives
a larger ratio as expected.}
\label{fig:sbg203} $ $ \vspace*{-1.2cm} \\
\end{figure}
\end{center}
\begin{center}
\begin{figure}[t] $ $ \vspace*{-1cm} \\
\centering
\epsfig{file=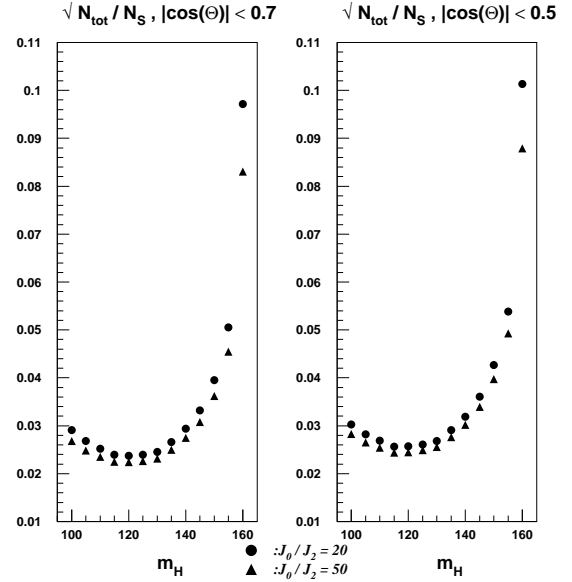,width=8.5cm} \vspace*{-1.3cm} \\
\caption{The statistical accuracy of the measurement based on a one year running
with the parameters of Fig. \ref{fig:ne203}. Surprisingly the larger thrust angle
cuts gives a slightly better statistical significance.}
\label{fig:sa203} $ $ \vspace*{-1.2cm} \\
\end{figure}
\end{center}
\vspace*{-1.7cm} 
In this section we are presenting first (conservative) results based on the
minimally achievable detector performance expected for the Tesla design \cite{bp}.
All relevant radiative corrections discussed above are included and the two jet
definition is the one discussed in Ref. \cite{ms3} based on Sterman-Weinberg 
parameters and identifying $\epsilon \sqrt{s} \equiv l_c$. We thus arrive at
\begin{equation}
\sigma_{\rm 2j} = \sigma^{DL}_{RG}(\epsilon\sqrt{s}) + \tilde\sigma_{\rm SV} +
\sigma_{\rm H}(\epsilon\sqrt{s},\epsilon,\delta) \; .
\label{eq:2jetbisbis}
\end{equation}
where the three contributions contain the RG-improved form factors, the exact
one loop subleading corrections and the hard Bremsstrahlung radiation inside the
cone of half-angle $\delta=20^o$ respectively. Fig. \ref{fig:ne203} assumes
a 70\% double b-tagging efficiency and a 3\% probability of counting a 
$c \overline{c}$-pair as a $b \overline{b}$-pair. The figures show two curves
for a ratio of $J_0/J_2=20$ and $50$ as well as differing cuts on the jet
thrust angle $\theta$. One can see clearly that a high charm suppression rate
is highly desirable. While the larger thrust angle cut $|\cos \theta | < 0.5$
leads to a larger ratio of S to BG events (see Fig. \ref{fig:sbg203}), the
inverse statistical significance, displayed in Fig. \ref{fig:sa203}, is slightly lower for
$|\cos \theta | < 0.7$. Although several questions concerning the dependence
of these Monte Carlo results on the various jet parameters should be addressed
more carefully, it seems safe to assume that statistically an ${\cal O} ( 2\%)$
measurement for the diphoton partial Higgs width is feasible\footnote{We assume
that BR$(H \longrightarrow b \overline{b})$ can be measured from Higgs-Strahlung at the 1\% level.}. More thorough
investigations will be presented in Ref. \cite{msk}.

\section{Conclusions}

In this paper we have included all available and relevant radiative corrections
for the measurement of the partial Higgs width $\Gamma (H \longrightarrow \gamma
\gamma)$. Conservative assumptions about the expected machine and detector
parameters lead to the preliminary result that over a four years of running
time the diphoton partial width can be measured with a statistical uncertainty
of ${\cal O} (2 \%)$. This level of precision could be very important for large pseudoscalar
Higgs masses in theories, such as the MSSM, with extended Higgs sectors and could
significantly enhance the kinematical reach for new physics.

\section*{Acknowledgments}

I would like to thank my collaborators W.J.~Stirling and V.A.~Khoze for their
contributions to the results presented here. In addition I am grateful for
interesting discussions with V.I.~Telnov, M.~Battaglia, G.~Jikia and S.~
Soldner-Rembold.

\end{document}